\documentclass[a4paper,11pt]{article}
\usepackage[left=2cm,right=2cm, top=2cm,bottom=2cm,bindingoffset=0cm]{geometry} 
\usepackage[utf8]{inputenc}
\usepackage{amsmath}
\usepackage{amssymb}
\usepackage{epsf}
\usepackage{cite}
\usepackage{graphicx}
\usepackage[english]{babel}
\usepackage{indentfirst}

\usepackage{xcolor}
\usepackage{color,soul}

\begin{document}
	
\begin{center}
	\textbf {Coherent radiation produced by a relativistic bunch of charged particles passing through a photonic crystal}	
\end{center}

\bigskip

\begin{center}
	\textbf{V.G. Baryshevsky}
\end{center}

%

\begin{abstract}
The equations, which describe generation of coherent radiation by a relativistic electron bunch moving in a photonic crystal, are obtained.  
Spectral-angular distribution of radiation and intensity of radiation in time-domain are derived.  
The interaction of the bunch electrons with the electromagnetic field induced by bunch itself at the initial  stage of producing coherent radiation  is shown contributing to the expression for the bunch current with the terms proportional to particle acceleration (proportional to interaction time and strength of electric field induced by bunch itself) along with the terms proportional to velocity. 
Induced radiation produced by the relativistic bunch is considered using the quantum physics approach.  	
\end{abstract}

\textbf
{\small{Keywords:}}
\textbf
{\small{relativistic bunch, induced radiation, crystal, photonic crystal}}	
	
\section{Introduction}
The microwave, optical and X-ray sources of spontaneous and induced radiation are used for variety of
applications: acceleration of charged particles, nuclear fusion research, space research 
industry, biology and medicine. High-power THz sources are very important devices to bring promising prospects to wide use. %
Accelerator based sources are studied by numerous authors through 
expansion to THz, optical and X-ray range of approaches and principles, 
which general serve for microwave sources 
\cite{THZ-review1,THZ-review2,Pierce2,Ch_7,Tsimring,Granatstein,PhysRev2019}.


Among spontaneous radiation mechanisms that underlie in generation of induced radiation let us mention diffraction and transition 
radiation, Smith-Parcell, parametric X-ray radiation (PXR) and magneto-bremsstrahlung radiation in ondulators.

General feature of the above listed radiation sources is the instability of an electron beam, which results in 
beam self-modulation and radiation of electromagnetic waves. The increment 
$\delta_0 \sim Im k_z$, where $k_z$ is the longitudinal wave number, describes the electron 
beam instability responsible for radiation process. Conventionally it
is determined by the unperturbed density $\rho_{b0}$ of electrons in the beam as follows:
$\delta_0 \sim \sqrt[3]{\rho_{b0}}$ 
%
\cite{Tsimring,Granatstein,Miller}. 
The threshold current density $j_{thr}$, which is
required for coherent electromagnetic oscillations to grow, in this case depends on the 
beam-wave interaction length $L$ as $j_{thr} \sim L^{-3}$.

Another law of electron beam instability was discovered in 
%
\cite{Ch_7,9,Ch_1,Ch_6}. 
It was found that for
an electron beam moving in space-periodic structure (crystal, photonic
crystal, spatially periodic slow wave structure (SWS)) for the case when Bragg diffraction 
could occur, the electron beam instability increment $\delta_0$ could turn out to be 
determined by $\delta_0 \sim \sqrt[4]{\rho_{b0}}$ rather than conventional law
$\delta_0 \sim \sqrt[3]{\rho_{b0}}$.

The law of electron beam instability inherent for the 
volume free electron laser (VFEL) 
%
\cite{Ch_7,PhysRev2019} 
is revealed when 
distributed feedback is formed by Bragg diffraction of the beam-induced electromagnetic
wave by a spatially periodic structure of a resonator. This law gives the very different dependence
of the threshold current density on the interaction length $L$:
$j_{thr} \sim L^{-3-2S}$, where $S$ is the number of additional waves arisen due to diffraction. 
When one additional wave arises then the threshold current density depends on the interaction length $L$ as follows:

\begin{equation}
\label{eq:A1}
j_{thr}\sim L^{-5}.
\end{equation}

Thus, this threshold current law differs from the usual law: $j_{thr}~\sim~L^{-3}$.
First lasing of VFEL, which uses the above described instability law, was presented in 
%
\cite{FirstLasing_NIM}. 
The
dispersion equation describing e-beam instability in the photonic crystal placed inside the
resonator was studied in 
%
\cite{PhysRev2019}.

Dispersion equation enables to find the increment of radiative instability of the electron beam. 


In present paper the equations, which describe generation of spontaneous and induced coherent radiation by a relativistic electron bunch moving in a photonic crystal, are obtained.  


The expressions for the spectral-angular distribution of radiation and intensity of radiation in time-domain are provided.  
%
At the initial  stage of producing coherent radiation 
the interaction of the bunch electrons with the electromagnetic field induced by bunch itself contributes to the expression for the bunch current with the terms proportional to particle acceleration (proportional to interaction time and strength of electric field induced by bunch itself) along with the terms proportional to velocity.   


Particularly, these terms are responsible for the radiative instability, which is described by the dispersion equation.


\section{Equations of radiation for a relativistic electron beam in photonic crystal}

To describe generation of induced radiation in a photonic crystal one should start from Maxwell equations:


\begin{eqnarray}
\label{eq1}
rot\vec {H}(\vec {r},t) = \frac{{1}}{{c}}\frac{{\partial \vec{D}(\vec {r},t)}}{{\partial t}} + 
\frac{{4\pi}}{{c}}\vec{j}(\vec{r},t),
~~rot\vec{E}(\vec{r},t) =- \frac{{1}}{{c}}\frac{{\partial \vec{H}(\vec {r},t)}}{{\partial t}}, \\
div\vec {D}(\vec {r},t) = 4\pi \rho (\vec {r},t) ,
~~\frac{{\partial \rho (\vec {r},t)}
}{{\partial t}} + div\vec {j} (\vec {r},t) = 0,\nonumber
\end{eqnarray}

\noindent where $\vec {E}(\vec {r},t)$ and  $\vec {H} (\vec
{r},t)$  are the strength of the electric and the magnetic field,
respectively;
$\vec {j} (\vec {r},t)$ and $\rho (\vec {r},t)$ are the current and
{charge} densities;
$D_{i} \left( {\vec
	{r},t^{\prime} } \right) = \int \varepsilon _{il} \left( {\vec
	{r},t , t^{\prime} } \right)E_{l} \left( {\vec {r},t^{\prime} }
\right)dt^{\prime} $,
%
%
where indices $i,l = 1,2,3$ correspond to $x,y,z$;
{in case if properties of the photonic crystal do not depend on
	time, its dielectric permittivity tensor reads as $\varepsilon
	_{il} (\vec {r},t-t^{\prime})= \frac{1}{2 \pi}
	\int_{-\infty}^{+\infty}\varepsilon _{il}\left( {\vec {r},\omega}
	\right) e^{-i\omega(t-t^{\prime})} d\omega$.}

The current and charge densities are defined as:

\begin{equation}
\label{eq2} \vec {j}\left( {\vec {r},t} \right) = e \sum_\alpha
{\vec {v}}_\alpha \delta(\vec {r} -\vec {r}_\alpha(t)),
{~\rho(\vec {r},t)=e \sum_\alpha \delta(\vec {r} -\vec
	{r}_\alpha(t)) = e \rho_b (\vec {r},t)\,,}
\end{equation}

\noindent where $e$ is the electron charge, $\rho _{b}(\vec
{r},t)$ is the beam density (the number of electrons per 1
cm$^{3}$).
The velocity ${\vec {v}}_{\alpha}={\vec {v}}_{\alpha} (t)$ of
electron with number $\alpha $
{can be obtained by relativistic equation of charge motion in an
	electromagnetic field (see section 17 in \cite{Landau1975})}:

\begin{equation}
\label{eq4} \frac{{d  {\vec {p}}_{\alpha} } }{{dt}} = m\frac{{d }
}{{dt}} (\gamma _{\alpha}  \vec{v}_{\alpha})= e\left\{ {\vec
	{E}\left( {  {\vec {r}}_{\alpha},t} \right) +
	\frac{{1}}{{c}}\left[ {  {\vec {v}}_{\alpha} \times \vec {H}\left(
		{  {\vec {r}}_{\alpha},t} \right)} \right]} \right\},
\end{equation}
\noindent where $\vec{p}_{\alpha}=\vec{p}_{\alpha}(t)$ is the
particle momentum, $\gamma _{\alpha}  = \left( {1 -
	{\textstyle{{v_{\alpha} ^{2}} \over {c^{2}}}}} \right)^{ -
	{\textstyle{{1} \over {2}}}}$ is the Lorentz factor, $\vec
{E}\left( { {\vec {r}}_{\alpha},t} \right)$ and $\vec {H}\left( {
	{\vec {r}}_{\alpha},t} \right)$ are the electric and magnetic
field strength at point $ {\vec {r}}_{\alpha} = \vec {r}_{\alpha}
(t)$, where the electron with number $\alpha $ is located.
{Following the exercise concluding section 17 in \cite{Landau1975} one can
	transform (\ref{eq4}) to equation for ${\vec {v}}_{\alpha}$ as
	follows:}
\begin{equation}
\label{eq3} \frac{{d  {\vec {v}}_{\alpha} } }{{dt}} =
\frac{{e}}{{m\gamma _{\alpha} } }\left\{ {\vec {E}\left( { {\vec
			{r}}_{\alpha} ,t} \right) + \frac{{1}}{{c}}\left[ {  {\vec
			{v}}_{\alpha}  \times \vec {H}\left( {  {\vec {r}}_{\alpha} ,t}
		\right)} \right] - \frac{{ {\vec {v}}_{\alpha} } }{{c^{2}}}\left(
	{  {\vec {v}}_{\alpha} \vec {E}\left( { {\vec {r}}_{\alpha} ,t}
		\right)} \right)} \right\},
\end{equation}

From equations (\ref{eq1}) one can obtain

\begin{equation}
\label{eq5}
- \Delta \vec {E}(\vec {r},t) + \vec {\nabla} \left( {\vec {\nabla} \vec {E}(\vec {r},t)} \right) +
\frac{{1}}{{c^{2}}}\frac{{\partial ^{2}\vec {D}(\vec
		{r},t)}}{{\partial t^{2}}} = - \frac{{4\pi}
}{{c^{2}}}\frac{{\partial \vec {j}(\vec {r},t)}}{{\partial t}}.
\end{equation}

In general case the dielectric permittivity tensor can be presented in the form:
$\hat{\varepsilon}(\vec{r})=1+\hat{\chi}(\vec{r})$, where $\hat{\chi}(\vec{r})$ is the susceptibility of a matter.
In a spatially-periodic structure (photonic crystal) susceptibility can be written as:

\begin{equation}
\label{eq:A2}
\hat{\chi}(\vec{r})=\sum_{\substack{i}} \hat{\chi}_{cell}(\vec{r}-\vec{r}_i),\nonumber
\end{equation}
where $\hat{\chi}_{cell}(\vec{r}-\vec{r}_i)$ is the susceptibility of the crystal unit cell. 
The susceptibility of a crystal can be expanded into a Fourier series as follows: 
$\hat{\chi}(\vec{r})=\sum_{\substack{\vec{\tau}}}\hat{\chi}_{\tau}e^{i\vec{\tau}\vec{r}}$,
where $\vec{\tau}$ is the reciprocal lattice vector of the crystal. Hereinafter it is supposed that components of susceptibility tensor $\hat{\chi}(\vec{r})$ comply the inequality  $\chi_{ik} \ll 1$.


From Maxwell equations (\ref{eq1}) it follows that:

\begin{equation} 
\label{eq:7}
div \vec{D}(\vec {r},\omega)= div \vec{E}(\vec {r},\omega) +
\sum_{ik}{\frac{\partial}{\partial x_{i}}}
 \left(
\chi_{ik}(\vec{r},\omega)E_{k}(\vec{r},\omega)
 \right)= 4\pi\rho (\vec {r},\omega),
\end{equation}

i.e.

\begin{equation}  
\label{eq:8}
div \vec{E} (\vec {r},\omega)= 4\pi\rho (\vec {r},\omega)
-\sum_{i,k}{\frac{\partial}{\partial x_{i}}} \left(  \chi_{ik} (\vec{r},\omega) E_{k} (\vec{r},\omega)
\right), 
\end{equation}

\noindent therefore,

\begin{equation}  
\label{eq:9}
\vec{\nabla}(\vec{\nabla}\vec{E} (\vec {r},\omega))=
4\pi\vec{\nabla}\rho (\vec {r},\omega) -
\vec{\nabla}\left(\sum_{i,k}{\frac{\partial}{\partial x_{i}}}
\left(\chi_{ik}(\vec{r},\omega)E_{k}(\vec{r},\omega)\right)\right).
\end{equation}

\noindent As a result equation (\ref{eq5}) can be rewritten as
follows:

\begin{equation}
\label{eq:10}
\Delta \vec{E}(\vec {r},t) + \vec{\nabla} 
\left(
\sum_{ik}  \int {\frac{\partial}{\partial x_{i}}} 
\left(
\chi_{ik}(\vec{r},t-t^{\prime})E_{k}(\vec{r},t-t^{\prime}) 
\right) dt^{\prime}
\right) -
\frac{1}{c^2}\frac{\partial^2}{\partial t^2} \int \hat{
	\varepsilon} (\vec{r}, t-t') \vec{E} (\vec{r}, t')dt^{\prime} = \frac{4
	\pi}{c^2} \frac{\partial \vec{j}(\vec {r},t)}{\partial t} + 4 \pi
\vec{\nabla} \rho (\vec {r},t). 
\end{equation}

The system of equations (\ref{eq1})-(\ref{eq:10}) 
enables studying radiation produced by an electron beam.
For further consideration let us make Fourier transform of (\ref{eq:10}): 

\begin{equation}
\label{eq:A5}
\vec{E}(\vec{r},t)=\frac{1}{2\pi}\int\vec{E}(\vec{r},\omega)e^{-i{\omega}t}d\omega,~
\vec{E}(\vec{r},\omega)=\int \vec{E}(\vec{r},t)e^{i{\omega}t}dt. \nonumber
\end{equation} 


\noindent Thus we obtain:

\begin{eqnarray}
\label{eq:11}
& & \Delta \vec{E}(\vec{r}, \omega) + \vec{\nabla}\left(
\frac{\partial}{x_i}\chi_{ik}(\vec{r},\omega) E_k(\vec{r}, \omega)\right)
+\frac{\omega^2}{c^2} \vec{E} (\vec{r}, \omega)
\nonumber
\\
& &+\frac{\omega^2}{c^2}\hat{\chi} (\vec{r})\vec{E}(\vec{r},
\omega)= -\frac{4 \pi i \omega}{c^2}\vec{j} (\vec{r}, \omega)+4
\pi\vec{\nabla} \rho (\vec{r}, \omega). 
\label{eq:AA1}
\end{eqnarray}
%
%
When describing wave diffraction in a spatially-periodic media (crystal) in case when components of susceptibility tensor $\chi_{ik} \ll 1$,  the second summand in (\ref{eq:AA1}) can be omitted \cite{PhysRev2019}.

Now we can consider spectral-angular distribution of the beam radiated energy $W_{\vec{n}\omega}$ per unit solid angle and analyse 
dependence of the intensity $I(t)$ of radiation, producing by beam which is passing through a generation range.

\section{Spectral-angular distribution of radiation produced by a bunch of relativistic particles}
\label{sec:3}

The spectral-angular density of radiation energy per unit solid angle $W_{\vec{n}\omega}$ produced by particles at a large distance 
$\vec{r}$ from the radiation region is known 
%
\cite{Landau1975}:

\begin{equation}
\label{eq:12}
W_{\vec{n}\omega}=\frac{cr^2}{4\pi^2}|\vec{E}(\vec{r},\omega)|^2
\end{equation}

In order to obtain $\vec{E}(\vec{r},\omega)$, Maxwell equation describing the interaction of particles within the generation range 
should be solved.
Maxwell equation solution describing the field $\vec{E}(\vec{r},\omega)$ at a large distance from the generation range can be 
obtained using Green function $G$, which is a tensor (for detailed description of the Green function see, for example 
\cite{Jackson,Morse}).

According to \cite{Bar_NO}, Green function is expressed at $r\rightarrow\infty$ through the solution of 
homogeneous Maxwell equations $\vec{E}_n^{(-)}(\vec{r},\omega)$
containing divergent spherical waves:

\begin{equation}
\label{eq:A13}
limG_{nl}(\vec{r},\vec{r\,}^{\prime},\omega)=\frac{e^{ikr}}{r}\sum_{\substack{s}}e^s_nE^{(-)s*}_{\vec{k}l}(\vec{r\,}^{\prime},\omega),
\end{equation}
where $n, l=x,y,z$, $e^s_n$ are components of the unit polarization vector
${\vec{e}}^s, s=1,2,\\
\vec{e\,}^1\perp\vec{e\,}^2\perp\vec{k},
\vec{k}=\frac{\omega}{c}\frac{\vec{r}}{r}$.

At a large distance from the generation range $r'\rightarrow\infty$ solution $\vec{E}^{(-)s}_k$
contains convergent spherical wave  and can be as follows:

\begin{equation}
\vec{E}^{(-)s}_{\vec{k}}(\vec{r\,}^{\prime},\omega)=\vec{e\,}^s  \left(e^{i\vec{k}\vec{r\,}^{\prime}}+const\frac{e^{-ikr'}}{r'} \right).
\end{equation}

Waves as $\vec{E}^{(-)s*}_{\vec{k}}$ are closely connected with the well-known Maxwell equation solutions 
$\vec{E}^{(+)s}_{\vec{k}}$,  that contain divergent spherical wave 
%
\cite{Bar_NO}:

\begin{equation}
\label{eq:15}
\vec{E}^{(-)s*}_{\vec{k}}(\vec{r},\omega)=\vec{E}^{(+)s}_{-\vec{k}}(\vec{r},\omega),
\end{equation}

\begin{equation}
\label{eq:16}
\vec{E}^{(+)s}_{\vec{k}}=\vec{e}^s  \left( e^{i\vec{k}\vec{r}}+const\frac{e^{ikr}}{r} \right).
\end{equation}

Thus, the solution $\vec{E}^{(-)s}_{\vec{k}}$ containing the asymptotically convergent 
spherical wave is used for describing the radiation produced inside the radiation range. This 
is contrary to the scattering of the external electromagnetic wave from the same generation 
region, when the electromagnetic field $\vec{E}^{(+)s}_{\vec{k}}$ includes a divergent
spherical wave at an infinite distance (see (\ref{eq:16})).
Both these solutions are simply connected with each other by the equation (\ref{eq:15}), which 
corresponds to the well-known optics reciprocity theorem 
%
\cite{Born_Wolf 1975,Bar_NO}.

Using Green function (\ref{eq:12}) from (\ref{eq:11}) we can obtain:

\begin{equation}
\label{eq:17}
\vec{E}({\vec{r}},\omega)=\frac{e^{ikr}}{r}\frac{i\omega}{c^2}\sum_{\substack{s}}
\vec{e\,}^s\int\vec{E}^{(-)s*}_{\vec{k}}(\vec{r\,}^{\prime},\omega)\bigg[\vec{j}(\vec{r\,}^{\prime},\omega)+\frac{ic^2}{\omega}\vec{\nabla}
\rho(\vec{r\,}^{\prime},\omega)\bigg]d^3r',
\end{equation}
thus:

\begin{equation}
\label{eq:18}
\vec{E}({\vec{r}},\omega)=\frac{e^{ikr}}{r}\frac{i\omega}{c^2}\sum_{\substack{s}}\vec{e\,}^s
\int\vec{E}^{(+)s}_{-{\vec{k}}}(\vec{r\,}^{\prime},\omega)
\bigg[\vec{j}(\vec{r\,}^{\prime},\omega)+\frac{ic^2}{\omega}\vec{\nabla}\rho(\vec{r\,}^{\prime},\omega)\bigg]d^3r',
\end{equation}
where $\vec{E}^{(+)s}_{-{\vec{k}}}$ is a solution of the homogenous equation (\ref{eq:11}), when right part of the equation 
is equal to zero.

As a result, the spectral-angular density of radiation energy of photons with polarizations
$W^s_{\vec{n}\omega}$ can be written in the following form:

\begin{equation}
\label{eq:19}
W^s_{\vec{n}\omega}=\frac{\omega^2}{4\pi^2c^3}\bigg\vert\int\vec{E}^{s(-)*}_{\vec{k}}(\vec{r},\omega)
[\vec{j}(\vec{r},\omega)+\frac{ic^2}{\omega}\vec{\nabla}\rho(\vec{r\,}^{\prime},\omega)]d^3r\bigg\vert^2,
\end{equation}
or
\begin{equation}
\label{eq:20}
W^s_{\vec{n}\omega}=\frac{\omega^2}{4\pi^2c^3}\bigg\vert\int\vec{E}^{(+)s}_{-{\vec{k}}}(\vec{r},
\omega)[\vec{j}(\vec{r},\omega)+\frac{ic^2}{\omega}\vec{\nabla}\rho(\vec{r\,}^{\prime},\omega)]d^3r\bigg\vert^2.
\end{equation}

Number of the radiated quants:

\begin{equation}
\label{eq:21}
dN^s_{\vec{n}\omega}=\frac{1}{\hbar\omega}W^s_{\vec{n}\omega}.
\end{equation}

\section{Time dependence of the intensity of radiation produced by a relativistic bunch}
\label{sec:4}

The intensity $I(t)$ of radiation produced at time $t$ by a relativistic electron bunch which is passing through a radiation range 
can be found with known electric field $\vec{E}(\vec{r},t)$, (magnetic field $\vec{H}(\vec{r},t)$) of the electromagnetic wave, which 
is produced by this bunch \cite{Landau1975}:

\begin{equation}
\label{eq:22}
I(t)=\frac{c}{4\pi}\mid\vec{E}(\vec{r},t)\mid^2r^2d\Omega
\end{equation}

Field:
 
\begin{equation}
\label{eq:222}
\vec{E}(\vec{r},t)=\frac{1}{2\pi}\int\vec{E}({\vec{r},\omega})e^{-i{\omega}t}d\omega.\nonumber
\end{equation}

Fourier transform $\vec{E}({\vec{r}},\omega)$ is determined by equations (\ref{eq:17}) and (\ref{eq:18}).

Using (\ref{eq:18}) for $\vec{E}(\vec{r},t)=\frac{1}{2\pi}\int\vec{E}(\vec{r},\omega)e^{-i{\omega}t}d\omega$ we can obtain 
following equation:

\begin{equation}
\label{eq:23}
\vec{E}(\vec{r},t)=-\frac{1}{r}\sum_{\substack{s}}\vec{e\,}^s\int\vec{E}^{(+)s}(\vec{r\,}^{\prime},t-\frac{r}{c}-t')
\bigg[\frac{1}{c^2}\frac{\partial\vec{j}(\vec{r\,}^{\prime},t')}{{\partial}t'}+\vec{\nabla}\rho(\vec{r\,}^{\prime},t')\bigg]d^3r'dt',
\end{equation}	
where
\begin{equation}
\label{eq:24}
\vec{E}^{(+)s}(\vec{r\,}^{\prime},t-\frac{r}{c}-t')=\frac{1}{2\pi}\int\vec{E}^{(+)s}_{-\vec{k}}(\vec{r\,}^{\prime},\omega)
e^{-i\omega(t-\frac{r}{c}-t')}d\omega.
\end{equation}

Thus $\vec{E}(\vec{r\,}^{\prime},t)$ can be rewritten as:

\begin{equation}
\label{eq:25}
\vec{E}(\vec{r},t)=\frac{1}{r}\vec{F}(t-\frac{r}{c}),
\end{equation}
where $\vec{F}(t-\frac{r}{c})$ is time dependent radiation amplitude.


Polarization $I_s(t)$ is defined 
by field $\vec{E}^s$, which is given by one of terms in the sum
 $\sum_{\substack{s}}$ in
(\ref{eq:23}). 
Though the obtained equations are general and exact, calculation of fields, $W^s_{\vec{n}\omega}$ and $I(t)$ could be quite difficult.
%
%
%
%
%
Current density and density $\rho$ in (\ref{eq:19}), (\ref{eq:20}), (\ref{eq:22})-(\ref{eq:25}) are given by sums of individual contributions from beam particles. 
Thus, radiation field is the sum of fields produced by each beam particle. 
%
%

Each beam particle interacts with electromagnetic field induced by itself and by all other beam particles, therefore each particle trajectory and velocity can barely be calculated exactly, thus making calculation of $W^s_{\vec{n}\omega}$ and $I(t)$ practically impossible. 
To find radiation field some approximations should be applied.
%

\section{Radiation in conditions when electromagnetic field weakly influence motion of beam particles} 

To begin with let us calculate spectral-angular distribution $W_{\vec{n}\omega}$ 
and intensity $I(t)$ in conditions when electromagnetic field weakly influence motion of beam particles.
%
%
Let us use approximation, which implies that radiated photons do not influence motion of bunch electrons.
%
%
%
In this case electron either moves with constant velocity or oscillate in the external field, which is produced by undulator or electromagnetic wave (dynamic undulator).
%
%
Obtained for this case $W_{\vec{n}\omega}$ and $I(t)$  describe spontaneous radiation produced by the bunch; these quantities can be expressed with the use of well-studied spectral-angular distribution $W^{(1)}_{\vec{n}\omega}$   and intensity $I^{(1)}(t)$ of spontaneous radiation produced by a single electron.
%
%
%
Let us now consider influence of radiated electromagnetic field on motion of electron bunch in the first-order perturbation theory and analyze the expression comprising microscopic beam current  in equation (\ref{eq:10}).
%
%

\begin{equation}
\label{eq:26}
\vec{J}{(\vec{r},t)}=4\pi\left[\frac{1}{c^2}\frac{\partial\vec{j}(\vec{r},t)}{{\partial}t}+
\vec{\nabla}\rho({\vec{r}},{t})\right],
\end{equation}
where $\vec{j}(\vec{r},t)=e\sum_{\substack{\alpha}}\vec{v}_\alpha(t)\delta(\vec{r}-\vec{r}_\alpha(t))$,
$\rho=e\sum_{\substack{\alpha}}\delta(\vec{r}-\vec{r}_\alpha(t))$.

Fourier-transform $\vec{J}(\vec{k},\omega)$ can be obtained by the use of (\ref{eq:10}) as follows:
%

\begin{equation}
\label{eq:27}
\vec{J}(\vec{k},\omega)=4\pi{\int}e^{-i\vec{k}\vec{r}}e^{i{\omega}t}\left[-\frac{i{\omega}e}{c^2}
\sum_{\substack{\alpha}} \vec{v}_\alpha(t)\delta(\vec{r}-\vec{r}_\alpha(t))+ie\vec{k}
\sum_{\substack{\alpha}}\delta(\vec{r}-\vec{r}_\alpha(t))\right]d^3rdt
\end{equation}

Integration over $d^3r$ and use of $\delta$ function definition give:
%

\begin{equation}
\label{eq:28}
\vec{J}(\vec{k},\omega)=4\pi{\int}e^{i{\omega}t}\left[-\frac{i{\omega}e}{c^2}\sum_{\substack{\alpha}}
\vec{v}_\alpha(t)e^{-i\vec{k}\vec{r}_\alpha(t)}
\theta(t-t_\alpha)
+ie\vec{k}\sum_{\substack{\alpha}}
e^{-i\vec{k}\vec{r}_\alpha(t)}\theta(t-t_\alpha) \right]dt
\end{equation}

Particle trajectory $\vec{r}_\alpha(t)$ can be expressed as follows:

\begin{equation}
\label{eq:29}
\vec{r}_\alpha(t)=\vec{r}_{{\alpha}0}+\vec{v}_\alpha(0)(t-t_\alpha)+\int_{t_\alpha}^{t}
\delta \vec{v}_\alpha(t')dt',
\end{equation}
where $\vec{r}_{\alpha0}$ are the initial  coordinates of particle $\alpha$,  $\vec{v}_\alpha(0)$ is the initial velocity of particle $\alpha$, 
$\delta\vec{v}_\alpha(t')$ is change of particle velocity caused by action of electromagnetic fields, $t_\alpha$ is the instant of particle emission by the electron beam source, $\theta(t-t_\alpha)=0$, when $t<t_\alpha$ and $\theta(t-t_\alpha)=1$ for $t \ge t_\alpha$. 


From (\ref{eq:28}) and  (\ref{eq:29}) it follows:

\begin{multline}
\label{eq:30}
\vec{J}(\vec{k},\omega)=4\pi\sum_{\substack{\alpha}}{\int}e^{i{\omega}t}\left[-\frac{i{\omega}e}{c^2}
\vec{v}_\alpha(0)+ie\vec{k}\right]e^{-i\vec{k}\left[\vec{r}_{{\alpha}0}+\vec{v}_\alpha(0)
(t-t_\alpha)+\int_{t_\alpha}^{t}\delta\vec{v}_\alpha(t')dt'\right]}dt+\\
+4\pi\sum_{\substack{\alpha}}{\int}e^{i{\omega}t}(-\frac{i{\omega}e}{c^2}\delta\vec{v}_\alpha
(t))e^{-i\vec{k}\vec{r}_\alpha(t)}dt.
\end{multline}

Let us now consider the initial stage of radiation generation, when phase acquired by a particle in a resonator (generation area, interaction area) $\vec{k}\int_{t_\alpha}^{t}{\delta}v_\alpha(t')dt'$ appears small as compared with unity: 
$\vec{k}\int_{t_\alpha}^{t}{\delta}v_\alpha(t')dt'\ll 1$ (the similar conditions take place if the length of resonator is small).
Radiation in indulator is not considered hereinafter. 
For this case  $J(k,\omega)$ reads:

\begin{multline}
\label{eq:31}
\vec{J}(\vec{k},\omega)=-4{\pi}i\sum_{\substack{\alpha}}{\int}e^{i{\omega}t}
\left[\frac{{\omega}e}{c^2}\vec{v}_{\alpha}(0)-e\vec{k}\right] 	
e^{-i\vec{k}\left[\vec{r}_{\alpha0}+\vec{v}_\alpha(0)(t-t_\alpha)\right]}
\bigg(1-i\vec{k}\int_{t_\alpha}^{t}\delta\vec{v}_\alpha(t')dt'\bigg)dt-\\
-\frac{4{\pi}i{\omega}e}{c^2}\sum_{\substack{\alpha}}{\int}e^{i{\omega}t}
\delta\vec{v}_\alpha(t)e^{-i\vec{k}\left[\vec{r}_{\alpha0}+\vec{v}_\alpha(0)(t-t_\alpha)\right]}dt.
\end{multline}

Current $\vec{J}(\vec{k},\omega)$ includes terms, proportional to $\delta\vec{v}_\alpha$; these terms describe contributions to
$\vec{J}(\vec{k},\omega)$, which are caused by influence of electromagnetic field on particle motion. 

%
Prior entering the interaction area  $\delta\vec{v}_\alpha=0$, when prticle leaves the interaction area, where field presents, change of velocity due to interaction with field is $\delta\vec{v}_\alpha=const$.
%
%
The term in (\ref{eq:31}), which does not include $\delta\vec{v}_\alpha$, describes  Fourier transform of current $\vec{J}_0(\vec{k},\omega)$ for pacticles moving with constant velocity:
%

\begin{equation}
\label{eq:32}
\vec{J}_0(\vec{k},\omega)=-4{\pi}i\sum_{\substack{\alpha}} \int e^{i{\omega}t}\left[\frac{{\omega}e}{c^2}
\vec{v}_\alpha(0)-e\vec{k}\right]e^{-i\vec{k}[\vec{r}_{\alpha0}+\vec{v}_\alpha(0)(t-t_\alpha)]}dt.
\end{equation}

Current $\vec{J}_0(\vec{k},\omega)$ is responsible for conventional spontaneous radiation (Smith-Purcell 
effect, diffarction radiation, Cherenkov radiation, transition radiation, parametric X-ray radiation, parametric (quasi-Cherenkov) radiation in a photonic crystal).
%
%

Let us analyze integrals over time, which present in terms proportional to   $\delta\vec{v}_\alpha$:
%

\begin{equation}
\label{eq:33}
I_1={\int}dt e^{i({\omega}-\vec{k}\vec{v}_\alpha(0)t)}(\int_{t_\alpha}^{t}\delta\vec{v}_\alpha(t')dt'),
\end{equation}

\begin{equation}
\label{eq:34}
I_2={\int}dt e^{i({\omega}-k\vec{v}_\alpha(0))t}\delta\vec{v}_\alpha(t)dt.
\end{equation}

Integration of (\ref{eq:33}) and (\ref{eq:34}) by parts gives:


\begin{equation}
\label{eq:35}
I_1={\int}dt\frac{e^{i({\omega}-\vec{k}\vec{v}_\alpha(0))t}}{(\omega-\vec{k}\vec{v}_\alpha(0))^2}
\frac{d\delta\vec{v}_\alpha(t)}{dt}dt,
\end{equation}

\begin{equation}
\label{eq:36}
I_2=\int\frac{e^{i({\omega}-\vec{k}\vec{v}_\alpha(0))t}}{i(\omega-\vec{k}\vec{v}_\alpha(0))}\frac{d\delta\vec{v}_\alpha(t)}{dt}dt.
\end{equation}

Therefore, $\vec{J}_0(\vec{k},\omega)$ can be expressed as follows:
%

\begin{multline}
\label{eq:37}
\vec{J}(\vec{k},\omega)=J_0(\vec{k},\omega)- 
4\pi\sum_{\substack{\alpha}}\left[\frac{{\omega}e}{c^2}
\vec{v}_{\alpha(0)}-e\vec{k}\right] 
e^{-i\vec{k}(\vec{r}_{\alpha0}-\vec{v}_\alpha(0)t_\alpha)}{\int}dt\frac{e^{i(\omega-\vec{k}\vec{v}_\alpha(0))t}}
{(\omega-\vec{k}\vec{v}_\alpha(0))^2}
\left(\vec{k}\frac{d\vec{v}_\alpha(t)}{dt}\right)dt+\\
+\frac{4{\pi\omega}e}{c^2}\sum_{\substack{\alpha}}
e^{-i\vec{k}(\vec{r}_{\alpha0}-\vec{v}_\alpha(0)t_\alpha)}\int\frac{e^{i(\omega-\vec{k}\vec{v}_\alpha(0))}t}{(\omega-\vec{k}
\vec{v}_\alpha(0))}\frac{d\vec{v}_\alpha(t)}{dt}dt
\end{multline}

According to motion equations  
(see 
\cite{Ch_7,Tsimring}), 
derivative 
$\frac{d\delta\vec{v}_\alpha(t)}{dt}=\frac{d\vec{v}_\alpha(t)}{dt}$ is determined by strength of electromagnetic field in the point, where particle is located. 

Therefore \eqref{eq:37} describes current expansion in series over power of  strength of field affecting the particle 
Integral ${\int}dte^{i(\omega-\vec{k}\vec{v}_\alpha(0))t}\frac{d\vec{v}_\alpha(t)}{dt}$ is proportional to Fourier-transform of the fields as follows:  $\vec{E}(\omega-\vec{k}\vec{v}_\alpha(0))$ and
$\vec{H}(\omega-\vec{k}\vec{v}_\alpha(0))$.


The term in (\ref{eq:37}), which is proportional to $\frac{1}{(\omega-\vec{k} v_\alpha(0))^2}$, is responsible for radiative instability of the electron beam  
(see, for example, 
%
\cite{Pierce2,Morse,Born_Wolf 1975}).
Let us consider how this term contributes to change of beam current with time.
Inverse Fourier transform of (\ref{eq:37}) ove frequency gives:


\begin{equation}
\label{eq:38}
\vec{J}(\vec{k},t)=\frac{1}{2\pi}\int_{-\infty}^{+\infty}\vec{J}(\vec{k},\omega)e^{-i{\omega}t}d\omega
\end{equation}

To calculate $\vec{J}(\vec{k},t)$ given by (\ref{eq:38}) with the use of (\ref{eq:37}) one has to take several integrals. Let us consider them as follows:

%
%
\begin{equation}
\label{eq:39}
S_1=\frac{1}{2\pi}\int_{-\infty}^{+\infty}\frac{e^{-i\omega(t-t')}}{(\omega-\vec{k}\vec{v}_\alpha(0))^2}d\omega=
\frac{\partial}{\partial\Omega}\frac{1}{2\pi}{\int}\frac{e^{-i\omega(t-t')}}{\omega-\Omega}d\omega,
\end{equation}
where $\Omega=\vec{k}\vec{v}_\alpha(0)$.

Integral (\ref{eq:39}) is to be calculated by the approach given by the theory of functions of complex variable. 
Path of integration is placed to make integral nonzero at   $t>t'$ to comply with the causality principle that finally gives:


\begin{equation}
\label{eq:40}
S_1=(t-t')e^{-i\vec{k}\vec{v}_\alpha(0)(t-t')}\theta(t-t'),
\end{equation}
where $\theta(t-t')$ is the unit Heaviside function: $\theta(t-t')=1$ for $t{\ge}t'$, and $\theta(t-t')=0$ for $t < t'$.
%

Let us now consider another integral

\begin{equation}
\label{eq:41}
S_2=\frac{1}{2\pi}\int_{-\infty}^{+\infty}\frac{{\omega}e^{-i\omega(t-t')}}{(\omega
-\Omega)^2}d\omega=i\frac{\partial}{{\partial}t}\frac{1}{2\pi}\int_{-\infty}^{+\infty}
\frac{e^{-i\omega(t-t')}}{(\omega-\Omega)^2}d\omega,
\end{equation}
which reads:

\begin{equation}
\label{eq:42}
S_2=\left[i+(t-t')\Omega\right]e^{-i\Omega(t-t')}\theta(t-t'),
\end{equation}
where $\Omega=\vec{k}\vec{v}_\alpha(0).$

Using (\ref{eq:40}) and (\ref{eq:42}) one can obtain the following expression for 
$\vec{J}(\vec{k},t)$:

\begin{multline}
\label{eq:43}
\vec{J}(\vec{k},t)=\vec{J}_0(\vec{k},t)+\sum_{\substack{\alpha}}\bigg[-\frac{4{\pi}ie\vec{v}_\alpha(0)}
{c^2}(\vec{k}\delta\vec{v}_\alpha(t))-\frac{4{\pi}e\vec{v}_\alpha(0)}{c^2}
(\vec{k}\vec{v}_\alpha(0))\int_{t_\alpha}^{t}(t-t')(\vec{k}\frac{d\vec{v}_\alpha(t')}{dt'})dt'+
\\+4{\pi}e\vec{k}\int_{t_\alpha}^{t}(t-t')(\vec{k}\frac{d\vec{v}_\alpha(t')}{dt'})dt'+
\frac{4{\pi}ie\delta\vec{v}_\alpha(t)}{c^2}(\vec{k}\vec{v}_\alpha(0))\bigg] 
e^{-i\vec{k}\vec{r}_{\alpha0}}e^{-i\vec{k}\vec{v}_\alpha(0)(t-t_\alpha)}.
\end{multline}

According to (\ref{eq:43}) interaction of a particle with a field at the initial radiation stage causes presence in the expression for current terms proportional to 
to particle acceleration $\frac{d\vec{v}_\alpha(t')}{dt'}$ (in accordance with motion equations
\cite{Tsimring}, which are proportional to the strength of electromagnetic field induced by beam particles).
%
In accordance with (\ref{eq:43}) in short time after start of radiation process   (at the initial  stage of producing coherent radiation) the expression for the bunch current includes terms proportional to squared time (particle acceleration is supposed to be constant).
In the considered approximation the electromagnetic field induced by beam particles is caused by spontaneous radiation induced by current $\vec{J}_0$.
These terms present in  (\ref{eq:26})-(\ref{eq:28})
along with conventional current $\vec{J}_0(\vec{k},t)$, which describes spontaneous radiation.

%
%

%

Expressions for current (\ref{eq:26})-(\ref{eq:28}) depend on particle coordinate  $\vec{r}_\alpha(t)$. 
If neglect the influence of radiated field, then particle coordinates are as follows 
$\vec{r}_\alpha(t)=\vec{r}_{\alpha0}+\vec{v}_\alpha(0)(t-t_\alpha)$ 
(this is valid for Cherenkov, quasi-Cherenkov and transition radiation in the absence of undulator). 
Therefore, exponent in (\ref{eq:28}) reads as follows: $e^{-i\vec{k}\vec{r}_\alpha(+)}=e^{-i\vec{k}(\vec{r}_{\alpha0}+\vec{v}_\alpha(0)
(t-t_\alpha))}.$ 

If influence of radiated field is considered, then particle coordinates are as follows:
$$\vec{r}_\alpha(t)=
\vec{r}_{\alpha0}+\vec{v}_\alpha(0)(t-t_\alpha)+\int_{t_\alpha}^{t}{\delta}\vec{v}_\alpha(t')dt'$$
(see also (\ref{eq:29})).

Thus exponent in (\ref{eq:30}) has the following form: 
$
e^{-i\vec{k}\left[\vec{r}_{{\alpha}0}+\vec{v}_\alpha(0)
	(t-t_\alpha)+\int_{t_\alpha}^{t}\delta\vec{v}_\alpha(t')dt'\right]}.
$
%

In a short time interval the change of velocity is $\delta\vec{v}_\alpha(t')=at'$, where 
$\vec{a}=\frac{d\vec{v}_\alpha(t')}{dt'}=\frac{d\delta\vec{v}_\alpha(t')}{dt'}$ is the particle acceleration.
Therefore, the phase, which determines exponent $e^{-i \phi}$, reads as follows
$\phi=\vec{k}
[\vec{r}_{\alpha0}+\vec{v}_\alpha(0)(t-t_\alpha)+\frac{1}{2}\vec{a}(t-t_\alpha)^2]$.

In contrast to the phase, which determines spontaneous radiation and is a linear function of time 
$\phi_{sp}=\vec{k}(\vec{r}_{\alpha0}+\vec{v}_\alpha(0)(t-t_\alpha))$, phase $\phi$ includes the term proportional to squared time. The corresponding correction presents in the Fourier transform of the current. 
Integral

\begin{equation}
\label{eq:44}
I_{sp}={\int}e^{i{\omega}t}e^{-i\vec{k}\vec{v}_\alpha(0)(t-t_\alpha)}\theta(t-t_\alpha)dt,
\end{equation}
converts to Gauss-type integral as follows:
\begin{equation}
\label{eq:45}
I={\int}e^{i{\omega}t}e^{-i\vec{k}\vec{v}_\alpha(0)(t-t_\alpha)}e^{-i\frac{1}{2}\vec{k}\vec{a}(t-t_\alpha)^2}\theta(t-t_\alpha)dt.
\end{equation}

Substitution of either  (\ref{eq:37}) or (\ref{eq:43}) to (\ref{eq:18}),~(\ref{eq:23}) enables to find all variables  $\vec{E}(\vec{r},\omega),\\
\vec{E}(\vec{r},t), W_{n\omega}^s$ and $I(t)$.

Using (\ref{eq:18}) and (\ref{eq:26}) field $\vec{E}(\vec{r},\omega)$ can be expressed as follows:
%

\begin{equation}
 \label{eq:46}
 \vec{E}(\vec{r},\omega)=-\frac{e^{ikr}}{4{\pi}r}\sum_{\substack{s}}\vec{e\,}^s\int\vec{E}^{(+)s}_{-\vec{k}}(\vec{r\,}^{\prime},\omega)
 \vec{J}(\vec{r\,}^{\prime},\omega)d^3r',
 \end{equation}
 
\noindent i.e.

\begin{equation}
\label{eq:47}
\vec{E}(\vec{r},\omega)=-\frac{e^{ikr}}{4{\pi}r}\sum_{\substack{s}}\vec{e\,}^s\frac{1}{(2\pi)^3}{\int}
\vec{E}^{(+)s}_{-\vec{k}}(\vec{r\,}^{\prime},\omega)e^{i\vec{q}\vec{r\,}^{\prime}}\vec{J}(\vec{q},\omega)d^3r'd^3q.
\end{equation}

Therefore, 
\begin{equation}
\label{eq:48}
\vec{E}(\vec{r},\omega)=-\frac{e^{ikr}}{4{\pi}r}\sum_{\substack{s}}\vec{e}\,^s{\int}\vec{E}^{(+)s}_{-\vec{k}}(-\vec{q},\omega)
\vec{J}(\vec{q},\omega)d^3q.
\end{equation}

When radiation occurs in a resonator with spatially periodic structure field  $\vec{E}^{(+)s}_{-\vec{k}}(\vec{r\,}^{\prime},\omega)$ is expressed as superposition of Bloch waves.
Fourier-transform $\vec{E}^{(+)s}_{-\vec{k}}(-\vec{q},\omega)$ in this case is a sum of delta-like functions with their peaks corresponding wave-vectors associated with   
$\vec{E}^{(+)s}_{-\vec{k}}(\vec{r\,}^{\prime},\omega)$. 
%

Therefore, radiation field (\ref{eq:48}) has the pattern similar to that arising due to diffraction of electromagnetic field in a crystal 
%
\cite{PhysRev2019,Bar_NO,Bar_Param}. 
%
This fact enables simplification of (\ref{eq:48}) to an approximate expression. Such simplification is reasonable when a particular radiation source should be described - we will not accomplish this simplification hereinafter since it requires computer simulation. 
%
%

\section{Lasing equations for a relativistic electron beam in a photonic crystal}


After start of generation process, those terms in (\ref{eq:43}), which grow with time, make the assumption of weak influence of radiated field inapplicable. 
Therefore, further consideration is focused on deriving the equations for the case when influence of radited field on beam electrons is significant.


Periodic structure is usually placed inside a waveguide. 
The following consideration is made for a straight waveguide of arbitrary cross-section uniform along waveguide length. Waveguide axis gives $z$ direction. 


Field inside the  waveguide can be represented by expansion over the waveguide eigenfunctions as follows:

\begin{equation}
\label{eq:49}
\vec{E}(\vec{r},t)=Re\sum_{\tau}A_{n}(z,t,k_{zn})e^{i(k_{zn}z-{\omega}t)}\vec{Y}_n(\vec{r}_\perp,k_{zn}).
\end{equation}

The waveguide eigenfunctions  $\vec{Y}_n(\vec{r}_\perp,k_{zn})$ describe field distribution over the waveguide cross-section, $k_{zn}=\sqrt{\frac{\omega^2}{c^2}-\varkappa^2_n}, \varkappa_n$  is the transverse wave number 
%
\cite{Jackson,Morse,Landau}, 
$\vec{r}_\perp$ 
are the coordinates in the plane orthogonal to axis  $z$.
Amplitude $A_n (z,t,k_{zn})$ is a function, which slowly changes at a period of time of about $T=\frac{2\pi}{\omega}$, which is typical for the selected mode family.


Analysis can be simplified by an assumption that components of susceptibility tensor $\chi_{ik}\ll 1$: this assumption enables to use the following expressions:


\begin{equation}
\label{eq:50}
\Delta\vec{E}(\vec{r},t)-\frac{1}{c^2}\frac{\partial^2\vec{E}}{{\partial}t^2}-\frac{1}{c^2}\frac{\partial^2}{{\partial}t^2}\int\hat{
\chi}(\vec{r},t-t')\vec{E}(\vec{r},t')dt'=4\pi
\left(\frac{1}{c^2}\frac{\partial\vec{j}(\vec{r},t)}{{\partial}t}+\vec{\nabla}\rho(\vec{r},t)\right).
\end{equation}

Let us substitute (\ref{eq:49}) to (\ref{eq:50}), multiply the result by  $\vec{Y^*}_n(\vec{r}_\perp,k_{zn})$ and integrate over
$d^2r_{\perp}$, thus obtain the following system of equations:


\begin{multline}
\label{eq:51}
2ik_{zn}\frac{{\partial}A_n(z,t,k_{zn})}{{\partial}z}+i\frac{2\omega}{c^2}\frac{{\partial}A_n}
{{\partial}t}+i\frac{1}{c^2}\frac{\partial}{\partial\omega}\left[\omega^2\sum_{\substack{n'}}
\int\vec{Y^*}_n(\vec{r}_\perp,k_{zn})\hat{\chi}(\vec{r})\vec{Y}_{n'}(\vec{r}_\perp,k_{zn'})
d^2r_\perp \right] \frac{{\partial}A_{n'}}{{\partial}t} +\\
+\frac{\omega^2}{c^2}\sum_{\substack{n'}}\int\vec{Y^*}_n
(\vec{r}_\perp,k_{zn})\hat{\chi}(\vec{r})\vec{Y}_{n'}(\vec{r}_\perp,k_{zn'})d^2r_{\perp}
A_{n'}(z,t,k_{zn})=\\
=8\pi\int\vec{Y^*}_n(\vec{r}_\perp,k_{zn})e^{-i(k_{zn}z-{\omega}t)}\left[e \frac{1}{c^2} 
\frac{\partial}{{\partial}t}\sum_{\substack{\alpha}}\vec{v}_\alpha\delta(\vec{r}-
\vec{r}_\alpha(t))+e\vec{\nabla}\sum_{\substack{\alpha}}\delta(\vec{r}-\vec{r}_\alpha(t))\right]
d^2r_\perp.
\end{multline}

Right part of (\ref{eq:51}), which is denoted as $g(z,t,\{\vec{v}_\alpha(0)\},\{\vec{r}_\alpha(0)\},\{t_\alpha\})$,  can be expressed as follows:

\begin{eqnarray}
	\label{eq:52}
& &	g(z,t,\{\vec{v}_\alpha(0)\},\{\vec{r}_\alpha(0)\},\{t_\alpha\})=\\
& &	=-\frac{8{\pi}i{\omega}e}{c^2}\int
	\vec{Y_n^*}(\vec{r}_\perp,k_{zn})\sum_{\substack{\alpha}}\vec{v}_\alpha(t)\delta(\vec{r}-\vec{r}_\alpha(t))e^{-i(k_{zn}z_\alpha(t)
		-{\omega}t)}
	\theta(t-t_\alpha)\theta(T_\alpha-t) d^2 r_{\perp}+ \nonumber	\\
& &	+8\pi\int\vec{Y_n^*}(\vec{r}_\perp,k_{zn})
	\sum_{\substack{\alpha}}\left[e\vec{\nabla}_\perp
	\delta(\vec{r}-\vec{r}_\alpha(t))+ik_{zn}\vec{i}_ze\delta(\vec{r}-\vec{r}_\alpha(t)) \right] e^{-i(k_{zn}z_\alpha(t)-{\omega}t)}\theta(t-t_\alpha)\theta(T_\alpha-t)d^2r_\perp= \nonumber \\
& &	=\sum_{\substack{\alpha}}g(z,t,\vec{v}_\alpha(0),\vec{r}_\alpha(0),t_\alpha), \nonumber
\end{eqnarray}
%
where $\{\vec{v}_\alpha(0)\}$ denotes the set of initial particle velocities, 
$\{\vec{r}_\alpha(0)\}$ is the set of initial particle coordinates, $\{t_\alpha\}$ is the set of instants associated with particle emission,  
$T_\alpha$ is the instant for $\alpha$-th particle escape from the interaction area,
function 
$g(z,t,\vec{v}_\alpha(0),\vec{r}_\alpha(0),t_\alpha)$ refers to $\alpha$-th particle of the beam and can be obtained by omitting the sum $\sum_{\substack{\alpha}}$ in $g(z,t,\vec{v}_\alpha(0),\vec{r}_\alpha(0),t_\alpha)$.

Expression (\ref{eq:52}) should be averaged over the initial distribution of particle velocities $v_\alpha(0)$, over initial coordinates for a particle entering the resonator
$\vec{r}_\alpha(0)$ and over the initial instants $t_\alpha$ for particles entering the resonator:

\begin{equation}
\label{eq:53}
g(z,t)=\left\langle g(z,t,\{\vec{v}_\alpha(0)\},\{\vec{r}_\alpha(0)\},\{t_\alpha\})\right\rangle,
\end{equation}
where $ \left\langle ...\right\rangle $ implies the averaging procedure over  $\vec{v}_\alpha(0)$, $\vec{r}_\alpha(0)$, $t_\alpha$.
The associated distribution function $w$ reads:

\begin{equation}
\label{eq:54}
w=w(\{\vec{v}_\alpha(0)\},\{\vec{r}_\alpha(0)\},\{t_\alpha\}).
\end{equation}

Let us consider expression  (\ref{eq:52}). 
Summation over $\alpha$ actually is equivalent to summation over the initial instants $t_\alpha$ for particles entering the resonator, over the initial particle velocities $v_\alpha(0)$ and initial coordinates for a particle entering the resonator $\vec{r}_\alpha(0)$:
%
%
$\sum_{\substack{\alpha}}=\sum_{\substack{t_\alpha}}\sum_{\substack{r_\alpha(0)}}\sum_{\substack{v_\alpha(0)}}.$

Let us analyze the sum over the initial instants $t_\alpha$ for particles entering the resonator. Averaging of this sum over the initial instants of particle entering into resonator results in integral as follows:

\begin{equation}
\label{eq:55}
\sum_{\substack{t_\alpha}}\rightarrow\int_{0}^{t}g(z,t,\vec{r}_\alpha(0),\vec{v}_\alpha(0),t_\alpha)
\dot{n}(t_\alpha,\vec{r}_{\perp\alpha}(0),\vec{v}_\alpha(0))dt_\alpha,
\end{equation}
where $\dot{n}(t_\alpha,\vec{r}_{\perp\alpha}(0),\vec{v}_\alpha(0))$ is the density of particle with transversal coordinates $\vec{r}_{\perp\alpha}(0)$ and velocity  $\vec{v}_\alpha(0)$ entering into resonator during time unit within interval   
$dt_\alpha$. 
%
Time instant $t_\alpha=0$ corresponds to the event of the very first particle 
If the resonator entering plane is located at $z=0$, then all the particles entering the resonator has the same $z$-coordinate and transversal coordinates $\vec{r}_{\perp\alpha}(0)$.
%
%

Integration over $dt_\alpha$ can be carried using delta-function in (\ref{eq:52}):
$\delta(\vec{r}-\vec{r}_\alpha(t))=\delta(\vec{r}_{\perp}-\vec{r}_{\perp\alpha}(t,t_\alpha))\delta(z-z_\alpha(t,t_\alpha))$.
Let us consider $\delta(z-z_\alpha(t,t_\alpha))$ remembering that


\begin{equation}
\label{eq:56}
\delta(\varphi(x))=\sum_{\substack{f}}\frac{\delta(x-x_f)}{\left|\left(\frac{d
		\varphi}{dx}\right) _{x=x_f}\right|},
\end{equation}
where $x_f$ are the roots of equation $\varphi(x)=0$.

Therefore
\begin{equation}
\label{eq:57}
\delta(z-z_\alpha(t,t_\alpha))=\frac{\delta(t_\alpha-t_f)}{\left|\left(\frac{dz_\alpha(t,t_\alpha)}{dt_\alpha}\right)_{t_\alpha=t_f}
\right|}.
\end{equation}

From (\ref{eq:17}) it follows that for a particle entering resonator  at instant $t_f,$ it appears at point $z=z_\alpha(t,t_f)$ at instant $t$. 
Result of integration of (\ref{eq:53}) and (\ref{eq:55}) gives  
for function $g(z,t)$ (see (\ref{eq:53})) the following expression:

\begin{equation}
\label{eq:58}
g(z,t)={\ll}g(z,t,\{\vec{v}_\alpha(0)\},\{\vec{r}_{\alpha\perp}(0)\},t_f)\gg,
\end{equation}
where $\ll....\gg$ means averaging over $v_\alpha(0)$ and 
$\vec{r}_{\alpha\perp}(0)$,

\begin{multline}
\label{eq:59}
g(z,t,\{\vec{v}_\alpha(0)\},\{\vec{r}_{\alpha\perp}(0)\},t_f)=-\frac{8{\pi}i{\omega}e}
{c^2}\sum_{\substack{\alpha}}\int\vec{Y_n^*}(\vec{r}_\perp,k_{zn})
\vec{v}_\alpha(t)\delta(\vec{r}_\perp-\vec{r}_{\alpha\perp}(t,t_f)) \times \\
\times e^{-i(k_{zn}z_\alpha(t,t_f)-{\omega}t)}\theta(t-t_f)\theta(T_f-t_f)\dot{n}(t_f,r_{\perp\alpha}
(0),\vec{v}_\alpha(0))d^2r_\perp+8\pi\sum_{\substack{\alpha}}\int\vec{Y_n^*}(\vec{r}_\perp,k_{zn})\\
\left[e\vec{\nabla}_\perp\delta(\vec{r}_\perp-\vec{r}_{\alpha\perp}(t,t_f))+ik_{zn}\vec{i}_ze\delta
(\vec{r}_\perp-\vec{r}_{\alpha\perp}(t,t_f))\right] \times e^{-i(k_{zn}z_\alpha(t,t_f)-{\omega}t)}\\
\dot{n}(t_f,\vec{r}_{\perp\alpha}(0),\vec{v}_\alpha(0))\theta(t-t_f)\theta(T_f-t)d^2r_\perp.
\end{multline}

Summation over $\alpha$ in (\ref{eq:59}) means summation over particle velocities and transversal coordinates.
Function $g(z,t,\{v_\alpha(0)\},\{\vec{r}_\alpha(0)\},t_f)$ comprises exponent 
$e^{-i(k_{zn}z(t,t_f)-{\omega}t)}=e^{-i\theta}$. 
Let us analyze phase 
$\theta=k_{zn}z_\alpha(t,t_f)-{\omega}t$.
Particle coordinate reads as follows:
\begin{equation}
\label{eq:60}
z_\alpha(t,t_f)=\int_{t_\alpha}^{t}v_{{\alpha}z(t')dt'}.
\end{equation}

Particle velocity  $v_{{\alpha}z}(t')$ reads as 
$v_{{\alpha}z}(t')=v_{{\alpha}z}(t_f)+{\delta}v_{{\alpha}z}(t')$, 
where $v_{{\alpha}z}(t_f)$ is the velocity of a particle entering the resonator, 
${\delta}v_{{\alpha}z}(t')$ describes change of velocity due to radiation and interaction with radiated electromagnetic field.
%
Therefore, particle coordinate reads as follows::
\begin{equation}
\label{eq:61}
z_\alpha(t,t_f)=v_{{\alpha}z}(t_f)(t-t_f)+\int_{t_f}^{t}{\delta}v_{{\alpha}z}(t')dt'.
\end{equation}

Thus, phase $\theta$ is expressed as follows:
\begin{equation}
\label{eq:62}
\theta_{(t,t_f)}=(k_{zn}v_{{\alpha}f}(t_f)-\omega)t+k_{zn}\int_{t_f}^{t}{\delta}v_{{\alpha}z}(t')dt'-k_{zn}
v_{{\alpha}z}(t_f)t_f
\end{equation}
and exponent $e^{-i\theta(t,t_f)}$ reads as follows: 
%
\begin{equation}
\label{eq:63}
e^{-i\theta(t,t_f)}=e^{-i(k_{zn}v_{{\alpha}f}(t_f)-\omega)t}e^{-ik_{zn}\int_{t_f}^{t}{\delta}v_{{\alpha}z}(t')dt}e^{+ik_{zn}
v_{{\alpha}z}(t_f)t_f}.
\end{equation}

Instants $t_f$ are randomly distributed instants of particle entering into resonator. 
%
In many practically important cases the average difference ${\Delta}t$ in entering instants is shorter than the period of radiated electromagnetic waves
${\Delta}t\ll\frac{2\pi}{\omega}$. 
Therefore, expression (\ref{eq:59}) can be averaged over entering instants  $t_f$ within the period $T=\frac{2\pi}{\omega}$ of electromagnetic wave oscillations.
%
%
Thus, equations (\ref{eq:52}), which describe generation process, can be written as:

\begin{multline}
\label{eq:64}
2ik_{zn}\frac{{\partial}A_n(z,t,k_{zn})}{{\partial}z}+i\frac{2\omega}{c^2}\frac{{\partial}A_n}{{\partial}t}+i\frac{1}{c^2}
\frac{\partial}{\partial\omega}\left[\omega^2\sum_{\substack{n'}}\int\vec{Y_n^*}(\vec{r}\perp,k_{zn})
\hat{\chi}(\vec{r})\vec{Y_{n'}}(\vec{r}\perp,k_{zn'})d^2r_\perp\right]\frac{{\partial}A_{n'}}{{\partial}t}+\\
+\frac{\omega^2}{c^2}\sum_{\substack{n'}}\int\vec{Y_n^*}(\vec{r}_\perp,k_{zn})\hat{\chi}
(\vec{r})\vec{Y_{n'}}(\vec{r}_\perp,k_{zn'})d^2r_{\perp}A_{n'}(z,t,k_{zn'})=g(z,t),
\end{multline}
where the right part $g(z,t)$ (see (\ref{eq:58}),~(\ref{eq:59})) can be written as follows:

\begin{equation}
\label{eq:65}
g(z,t)=\ll\frac{1}{T}\int_{t}^{t+\frac{2\pi}{\omega}}g(z,t,\{\vec{v}_\alpha(0)\},\{\vec{r}_{\alpha\perp}(0)\},t_f)dt_f\gg.
\end{equation}

Mind that notation $\ll...\gg$ means averaging over initial velocities and initial coordinates $\vec{r}_{\alpha\perp}(0)$ for a particle entering the resonator in the plane orthogonal to axis $z$ (transversal plane). 
%
%
Distribution of particles in the transversal plane is nonuniform. 
Beam moving in resonator can have various shapes: annular, sheet or even comprise multiple beams. 
%
Moreovere, distribution of fields in the transversal plane is also nonuniform. 
%
%
Therefore phase $\theta$ (see (\ref{eq:62})) could depend on the transversal coordinate.

To solve equations (\ref{eq:64}),~(\ref{eq:65}), which describe evolution of electromagnetic field, one should use equation of motion for beam particles 
%
%
\cite{Tsimring}. 
Joint solution of equations (\ref{eq:64}),~(\ref{eq3}) makes it possible to analyze generation of induced radiation.
%
However, to obtain evolution of electromagnetic field  taking into account field influence on motion of particles one should  numerically solve system of equations (\ref{eq:64}) and (\ref{eq3}).

The above  system of equations enables to describe lasing in VFEL in case when bunch of relativistic particles moves in a photonic crystal: 
%
%
both fields in resonator and influence of these fields on electron beam current can be obtained. 
%
%
When electron beam current is defined one can describe the spectral-angular distribution $W_{\vec{n}\omega}$ and evolution of radiation intensity  $J(t)$ (see sections \ref{sec:3} and \ref{sec:4}).
%

\section{Induced radiation produced by a bunch: quantum description}

It is well-known that probabilities of emission of spontaneous and stimulated (induced) radiation are described by Einstein coefficients.
%
%
Therefore, they can be used to describe amplification of intensity of radiation emitted by beam particles 
%
%
\cite{Ginzburg_theor, Ginzburg_Tsyt, Fedorov, Marshall}. 

Applying quantum conservation laws: laws of momentum and energy conservation, one can calculate corrections for quantum recoil. 
These corrections cause shift of spectral-angular distributions associated with photon emission by a particle with respect to distributions associated with photon absorption by a particle.
%
Therefore, probabilities of spontaneous emission and absorption are also different for the emitted and absorbed photons, which frequencies are equal (this correspond to the equal angle between photon emission direction and initial momentum of the particle and that between momentum of absorbed photon  and initial momentum of the particle).
%

Quantum conditions for stimulated amplification of radiation in the particle beam have been studied for a beam moving in a boundless medium. 
%
%
Analysis presented hereinafter is given for emission and absorption of radiation produced  by a bunch for both cases: a boundless and bounded media.
%

%
Radiation types to be considered are Cherenkov radiation in a parallel-sided plate in the frequency range corresponding to visible light (in this range dielectric permittivity can be much greater than 1), parametric X-ray radiation in a crystal, quasi-cherenkov parametric radiation in a periodic metamaterial (photonic crystal).
%

Since correction for quantum recoil is small, then emission and absorption probability can be expanded over this correction as a small parameter. Thus, analysis of radiation amplification can be reduced to calculating the derivative of probability of spontaneous radiation and, then, averaging the result over distribution of particle momentum in the beam
%
%
 \cite{Ginzburg_theor, Ginzburg_Tsyt, Fedorov, Marshall}. 
 In the considered case it implies calculation of the derivative of probability of spontaneous radiation produced by the bunch and, then, averaging the result over distribution of particle momentum in the bunch.

Expressions, which describes spontaneous radiation produced by the bunch, are obtained in \cite{Bar_NO,Bar_Xray} for parametric X-ray radiation in a crystal (quasi-cherenkov parametric radiation, diffraction radiation of an oscillator).
%
%
There it was shown that probability of spontaneous radiation produced by the bunch is determined by the  probability of spontaneous radiation produced by a single particle and by the form-factor of the bunch.
%
%
The expressions for probability of spontaneous radiation produced by a single particle are derived in \cite{Bar_NO,Bar_Param}.

\subsection{Cherenkov radiation in uniform medium}

Analysis of induced radiation from a bunch requires consideration of conservation laws. Let us consider a bunch  moving in uniform medim. 
Energy and momentum conservation laws in case of emission (absorption) of a photon reads:
%

\begin{equation}
\label{eq:66}
E_0=E_1 \pm \hbar \omega ;
\vec{p_0}=\vec{p_1} \pm \hbar \vec{k},
\end{equation}
where $E_0, \vec{p_0}$ are the energy and momentum of a particle before emission (absorption) of a photon, $\hbar \vec{k}$ is the momentum of the photon; 
$ \hbar \vec{k}= \frac{\hbar \omega}{c} n_l(\omega, \vec{s}) \vec{s}$, $\omega$ is the frequency of the photon, 
$n_l(\omega, \vec{s})$ is the refraction index for the wave of type $l$, which propagates in the considered medium, $\vec{s}$ is the unit vector along the propagation direction, $c$ is the speed of light.

Mind that particle energy and moment relate as  $ E_{0,1} = \sqrt{p_{0,1}^2c^2 + m^2c^4} $, where $m$ is the particle mass.
%

Using (\ref{eq:66}) one can get the expression for angle $\vartheta_\gamma$ between direction of photon emission and initial momentum of the particle for photon frequency $\omega$ 
\cite{Ginzburg_theor}: 

\begin{equation}
\label{eq:67}
\cos \vartheta_{\gamma}^{rad}= \frac{c}{vn}\left[ 1 + \frac{\hbar \omega_{rad}}{2E} (n^2-1) \right],
\end{equation}

\begin{equation}
\label{eq:68}
\hbar \omega_{rad}= \frac{2E}{cn (1-\frac{1}{n^2})} (v \cos\vartheta_{\gamma}^{rad} - \frac{c}{n}).
\end{equation}

Using the similar approach one can obtain the expression for angle  $\vartheta_\gamma^{absorb}$ (between  momentum of photon and initial momentum of the particle) for the absorbed photon with frequency $\omega$ as follows:

\begin{equation}
\label{eq:69}
\cos \vartheta_{\gamma}^{absorb}= \frac{c}{vn}\left[ 1 - \frac{\hbar \omega_{absorb}}{2E} (n^2-1) \right],
\end{equation}

\begin{equation}
\label{eq:70}
\hbar \omega_{absorb}= \frac{2E}{cn (1-\frac{1}{n^2})} (v \cos\vartheta_{\gamma}^{absorb} - \frac{c}{n}).
\end{equation}

Note that for general Maxwell equations $\vartheta_{\gamma}^{rad}$ is defined by:

\begin{equation}
\label{eq:71}
\cos\vartheta_{\gamma}^{absorb}= \frac{c}{vn}.
\end{equation}

Therefore, consideration of  quantum recoil for photon emission described by  (\ref{eq:67}-\ref{eq:68}) results in small correction to expression (\ref{eq:71}). 
%
%
Photon absorption can be described similarly. 
From the above consideration it follows that angles
$\vartheta_{\gamma}^{rad}$ and $\vartheta_{\gamma}^{absorb}$ differ from each other, as well as frequencies $\omega_{rad}$ and $\omega_{absorb}$. 
Therefore, probability of radiation and probability of absorption reach maximal value at different conditions
\cite{Ginzburg_theor, Ginzburg_Tsyt, Fedorov, Marshall}.

\subsection{Cherenkov radiation in a parallel-sided plate}

Presence of boundaries causes for photon wavefunction conversion from the plane wave to superposition of waves, which momentum changes due to refraction by the boundary
%
\cite{Landau}. 
Therefore conservation laws read as follows:

\begin{equation}
	\label{eq:72}
	E_0=E_1 \pm \hbar \omega, 
	p_{0z}=p_{1z} \pm \hbar {k_z},\vec{p}_{0\perp}=\vec{p}_{1\perp} \pm \hbar \vec{k}_{0 \perp},
\end{equation}

where $k_z=\sqrt{k_0^2 \varepsilon - k_{0 \perp}^2}$, $ k_0=\frac{\omega}{c}$, axis $z$ is orthogonal plate plane, $(x,y)$ plane coincides with the plate surface, $k_{0 \perp}^2=k_{0x}^2+k_{0y}^2$,
 $k_{0 \perp}^2=k_{0}^2 \sin^2 \vartheta$, $\vartheta$ is the between $\vec{k}_0$ and unit vector $\vec{s}_z$ (directed along $z$) in vacuum. 
 
Vector component $k_z$ of the wavevector $\vec{k}$ reads as follows:
\begin{equation}
 \label{} 
k_z = \frac{\omega}{c} \sqrt{\varepsilon - \frac{k_{0 \perp}^2}{k_{0}^2}}=\frac{\omega}{c}\sqrt{\varepsilon - sin^2 \vartheta}.
\nonumber
 \end{equation}

Let us analyze the following expression:
\begin{equation}
\label{eq:73}
p_{0z} - p_{1z} \mp \hbar {k_z}=0
\end{equation}
using $p_{0z}= p_0 \cos \vartheta_0, p_{1z}= p_1 \cos \vartheta_1$, $\vartheta_0 (\vartheta_1)$ is the angle between  
$\vec{p}_{0z} (\vec{p}_{1z})$ and $\vec{s}_z$.

Let us assume that $ \vartheta_0, \vartheta_1 \ll 1 $, i.e. incident particle enters the plane approximately normally. 
Recall that:
\begin{equation}
\label{}\nonumber
p_0 = \frac{1}{c} \sqrt{E^2 -m^2c^4},p_1 = \frac{1}{c} \sqrt{E_1^2 -m^2c^4}=\frac{1}{c} \sqrt{(E \mp \hbar \omega)^2 -m^2c^4}.
\end{equation}

Suppose that particle energy $E$ is greater than photon energy $\hbar \omega$  and use $\hbar \omega \ll E$ to obtain the following expression:
\begin{equation}
\label{eq:74*}
p_1 = \frac{1}{c} \sqrt{E^2 \mp 2E \hbar \omega +\hbar^2 \omega^2 - m^2 c^4} = p_0 \sqrt{1 \mp \frac{2E \hbar \omega}{p_{0}^{2}c^2} +\frac{\hbar^2 \omega^2}{p_{0}^{2}c^2}} , \nonumber
\end{equation}
i.e.
\begin{equation}
\label{eq:74}
p_1 \simeq p_0 \left( 1 \mp \frac{2E \hbar \omega}{p_{0}^{2}c^2} + \frac{\hbar^2 \omega^2}{2p_{0}^{2}c^2} - \frac{1}{8} \frac{4 E^2 \hbar^2 \omega^2}{p_{0}^{4}c^4}\right).
\end{equation}

Component $p_{1z}$ reads:

\begin{eqnarray}
\label{eq:75}
&&p_{1z} = p_1 \sqrt{1 - \frac{p^2_{1 \perp}}{p^2_{\perp}}} \simeq p_1 \left( 1 - \frac{p^2_{1 \perp}}{2 p^2_{1}} \right) = 
p_1 \left(1 - \frac{p^2_{0 \perp} \mp 2 \vec{p}_{0 \perp}\hbar \vec{k}_{0 \perp} +\hbar^2 k^2_{\perp}}{2 p_1^2} \right), \\ \nonumber
&&p_1^2 = p_0^2 \mp \frac{2E \hbar \omega}{c^2}+\frac{\hbar^2 \omega^2}{c^2}.
\end{eqnarray}

If the initial momentum of the particle is directed along  $ z $ (normal incidence), then $  \vec{p}_{0 \perp} = 0 $. 
Therefore:
\begin{eqnarray}
\label{eq:76}
&&p_{1z} = p_1 - \frac{\hbar^2 k^2_{\perp}}{2p_1} \simeq p_0 \left(1 \mp \frac{E \hbar \omega}{p_{0}^{2}c^2} + \frac{\hbar^2 \omega^2}{2p_{0}^{2}c^2} -\frac{1}{2} \frac{E^2 \hbar^2 \omega^2}{p_{0}^{4}c^4} \right) - \frac{\hbar^2 k^2_{ 0 \perp}}{2p_0}= \\ \nonumber
&&= p_0 \mp \frac{\hbar \omega}{v} + \frac{\hbar^2 \omega^2}{2p_{0}c^2} - \frac{1}{2} \frac{\hbar^2 \omega^2}{p_0 v^2} - \frac{\hbar^2 k^2_{\perp}}{2 p_0}= \\ \nonumber
&&= p_0 \mp \frac{\hbar \omega}{v} - \frac{\hbar^2 \omega^2}{2p_0 v^2} \left(1 - \frac{v^2}{c^2} \cos^2 \vartheta \right) .
\end{eqnarray}

Therefore (\ref{eq:73}) can be presented as follows:

\begin{eqnarray}
\label{eq:77*}
\pm \frac{\hbar \omega}{v} + \frac{\hbar^2 \omega^2}{2p_{0}v^2} \left( 1 - \frac{v^2}{c^2} \cos^2 \vartheta \right) \mp \frac{\hbar \omega}{c} \sqrt{\varepsilon - \sin^2 \vartheta}=0, \nonumber
\end{eqnarray}

i.e.

\begin{eqnarray}
\label{eq:77}
\pm \left( 1- \frac{v}{c} \sqrt{\varepsilon - \sin^2 \vartheta}\right) + \frac{\hbar \omega}{2p_0v} \left( 1- \frac{v^2}{c^2} \cos^2 \vartheta  \right) = 0.
\end{eqnarray}

Vavilov-Cherenkov condition for a plate reads:
%
$ 1- \frac{v}{c} \sqrt{\varepsilon- \sin^2 \vartheta}=0 $.
Therefore, 
$\frac{v^2}{c^2} (\varepsilon - \sin^2 \vartheta) =1$, 
i.e. $\cos^2 \vartheta + (n^2-1) =\frac{c^2}{v^2}$.
Therefore,  $ 1-\frac{v^2}{c^2}\cos^2 \vartheta = \frac{v^2}{c^2} (n^2-1) $
and (\ref{eq:77}) gives for radiation and absorption the following: 

\begin{eqnarray}
\label{eq:78}
\text{for emission~~~} 1-\frac{v}{c} \sqrt{\varepsilon - \sin^2 \vartheta} + \frac{v~\hbar \omega (n^2-1)}{2p_0c^2}=0 , \\ \nonumber
\text{for absorption~~~} 1-\frac{v}{c} \sqrt{\varepsilon - \sin^2 \vartheta} - \frac{v~\hbar \omega  (n^2-1)}{2p_0c^2}=0.
\end{eqnarray}

Calculating frequency $\omega$ from (\ref{eq:78}) one can get frequency shift due to quantum recoil effects.
%
%
For relativistic case $ \frac{v}{c} $ in the third summand of (\ref{eq:78}) can be set equal to $1$. 
%

When studying quantum recoil effect for quasi-cherenkov radiation one should replace in (\ref{eq:72}) momentum $\hbar 
\vec{k}  $ by the photon momentum in a spatially periodic medium (i.e. by photon wave vector in a spatially periodic medium $ \hbar \vec{k}_l $), the expressions for  $ \hbar \vec{k}_l $ are obtained in 
%
\cite{Bar_NO, Bar_Param}).

\subsection{Transition radiation}

Let us now consider correction for quantum recoil in case of  transition radiation. In this case in contrast to Vavilov-Cherenkov radiation the conservation laws reads as follows: 
%

\begin{equation}
\label{eq:79}
E_0=E1 \pm \hbar \omega, ~~ \vec{p}_0 =\vec{p}_1 \pm \hbar \vec{k} + \vec{q},
\end{equation}
where $\vec{q}$ is the transmitted momentum, $ \vec{k} = \frac{\omega}{c} \vec{s}_{\gamma} $.
Note that expressions (\ref{eq:79}) are also valid for  bremsstrahlung.
%

According to (\ref{eq:79}) the transmitted momentum can be expressed as follows: 

\begin{equation}
\label{eq:80}
\vec{q}=\vec{p}_0-\vec{p}_1 \mp \hbar \vec{k}.
\end{equation}

Multiplying $\vec{q}$ by unit vector $\vec{s}$ directed along the initial particle momentum $\vec{p}_0={p}_0\vec{s}$.

\begin{equation}
\label{eq:81}
{q}_{\parallel}={p}_0-{p}_1 \cos \vartheta_1 \mp \hbar \frac{\omega}{c} \cos \vartheta_{\gamma},
\end{equation}
where $\vartheta_{1}$ is the angle between momenta $ \vec{p}_0 $ and $ \vec{p}_1 $, $ \vartheta_{\gamma} $ is the angle between momenta  $\vec{p}_0$ and $\vec{k}$.
 
Typically momentum $\hbar \vec{k}$ of a photon produced by transition radiation  is  smaller than initial particle momentum  $\vec{p}_0$. Therefore $\vartheta_{1} \ll 1$ and the following expression is valid:
 
 \begin{equation}
 \label{eq:82}
 {q}_{\parallel}={p}_0 - {p}_1 - {p}_1 \frac{\vartheta_1^2}{2} \mp \frac{\hbar\omega}{c} \cos \vartheta_{\gamma}.
 \end{equation}

Momentum ${p}_1$ is expressed as follows (see (\ref{eq:74})):

\begin{equation}
\label{eq:82*}
p_1= \frac{1}{c}\sqrt{(E \mp \hbar \omega)^2 -m^2c^4}= p_0\left( 1 \mp \frac{E \hbar \omega}{p_0^2c^2} + \frac{\hbar^2 \omega^2}{2p_0^2c^2} - \frac{1}{2} \frac{E^2 \hbar^2 \omega^2}{p_0^4c^4}\right). \nonumber
\end{equation}

Projection $p_1 \cos \vartheta_1 $ is expressed by (\ref{eq:76}) ($z$ is directed along $ \vec{p}_0 $). Therefore:

\begin{equation}	
\label{eq:83}
 {q}_{\parallel}= \pm \frac{\hbar \omega}{v} + \frac{\hbar^2 \omega^2}{2p_0v^2} \left( 1- \frac{v^2}{c^2} \cos^2 \vartheta_{\gamma}\right) \mp \hbar \frac{\omega}{c} \cos \vartheta_{\gamma},
\end{equation}
i.e.
\begin{equation}	
\label{eq:84}
{q}_{\parallel}= \frac{\hbar \omega}{v}\left(\pm (1- \frac{v}{c} \cos \vartheta_{\gamma}) +  \frac{\hbar \omega}{2p_0v} (1-\frac{v^2}{c^2} \cos^2 \vartheta_{\gamma})\right).
\end{equation}

\begin{equation}	
\label{eq:85}
|q|_{\parallel}= \frac{\hbar \omega}{v}\left( (1- \frac{v}{c} \cos \vartheta_{\gamma}) \pm \frac{\hbar \omega}{2p_0v} (1-\frac{v^2}{c^2} \cos^2 \vartheta_{\gamma}) \right).
\end{equation}

According to (\ref{eq:84}, \ref{eq:85}) the quantum correction is included into the expression for the longitudinal incident momentum with different signs for emitted and for absorbed photons. This is similar to Cherenkov radiation case and thus, similarly results different conditions for probability of radiation and probability of absorption to reach their maximal values.  


Since correction is small, then emission and absorption probability cab be expanded over this correction as a small parameter. Thus, analysis of radiation amplification can be reduced to calculating the derivatives $ \Delta \vartheta_{\gamma} $ and $ \Delta \omega $ (i.e. $d\vartheta_{\gamma}$ and $d \omega$)  of probability of spontaneous radiation obtained in 
%
\cite{Bar_NO, Bar_Xray} and then
%
one should average the obtained expressions with the distribution of particle momentum
\cite{Ginzburg_theor, Ginzburg_Tsyt, Fedorov, Marshall}.  

\section{Conclusion}

In contrast to spontaneous radiation produced by a bunch, at the initial  stage of producing induced coherent radiation 
the interaction of the bunch electrons with the electromagnetic field induced by the bunch itself contributes to the expression for the bunch current with the terms proportional to particle acceleration (proportional to interaction time and strength of electric field induced by bunch itself) along with the terms proportional to velocity.   
%

The expressions for the spectral-angular distribution of radiation and intensity of coherent radiation in time-domain for a bunch  are provided.
%

Quantum description is given for the induced radiation of a bunch.
Probabilities of emission of spontaneous and stimulated (induced) radiation described by Einstein coefficients
enable to express amplification of intensity of radiation emitted by a bunch using 
probability of spontaneous radiation produced by the bunch, which is determined by the  probability of spontaneous radiation produced by a single particle and by the form-factor of the bunch.


\end{document}